\documentclass[sigconf]{acmart}

\AtBeginDocument{%
  \providecommand\BibTeX{{%
    \normalfont B\kern-0.5em{\scshape i\kern-0.25em b}\kern-0.8em\TeX}}}

\setcopyright{rightsretained}
\copyrightyear{2022}
\acmYear{2022}
\acmConference[WWW '22 Companion] {Companion Proceedings of the Web Conference 2022}{April 25--29, 2022}{Virtual Event, Lyon, France.}
\acmBooktitle{Companion Proceedings of the Web Conference 2022 (WWW '22 Companion), April 25--29, 2022, Virtual Event, Lyon, France}
\acmISBN{978-1-4503-9130-6/22/04}
\acmDOI{10.1145/3487553.3524712}

\acmSubmissionID{0148}

\usepackage{amsmath}
\usepackage{pifont}
\usepackage{cases}
\usepackage{algorithm}
\usepackage{listings}
\usepackage{multirow}
\usepackage{subfigure}
\usepackage{array}
\newcolumntype{C}[1]{>{\centering\let\newline\\\arraybackslash\hspace{0pt}}m{#1}}

\usepackage{etoolbox}
\makeatletter
\AfterEndEnvironment{algorithm}{\let\@algcomment\relax}
\AtEndEnvironment{algorithm}{\kern2pt\hrule\relax\vskip3pt\@algcomment}
\let\@algcomment\relax
\newcommand\algcomment[1]{\def\@algcomment{\footnotesize#1}}
\renewcommand\fs@ruled{\def\@fs@cfont{\bfseries}\let\@fs@capt\floatc@ruled
  \def\@fs@pre{\hrule height.8pt depth0pt \kern2pt}%
  \def\@fs@post{}%
  \def\@fs@mid{\kern2pt\hrule\kern2pt}%
  \let\@fs@iftopcapt\iftrue}
\makeatother

\usepackage{mathtools}
\usepackage{amsthm}
\usepackage{color}
\usepackage{multirow}

\newcommand{\ourname}{\textsc{LinkProp}}
\newcommand{\ournamemulti}{\textsc{LinkProp-Multi}}
\newcommand{\hy}{\hat{y}}
\newcommand{\hyu}[1]{\hat{y}(#1|u)}
\newcommand{\hyuone}[1]{\hat{y}(#1|u_{1})}
\newcommand{\hg}[1]{\hat{\Gamma}(#1)}

\begin{document}

\title{Revisiting Neighborhood-based Link Prediction\\ for Collaborative Filtering}


\author{Hao-Ming Fu}
\email{hfu2@andrew.cmu.edu}
\authornote{Work done during an internship at Snap Inc.}
\affiliation{
  \institution{Carnegie Mellon University \\Snap Inc.}
  \country{United States of America}
}

\author{Patrick Poirson}
\email{ppoirson@snapchat.com}
\authornote{Both authors contributed equally to this research.}
\affiliation{
  \institution{Snap Inc.}
  \country{United States of America}
}

\author{Kwot Sin Lee}
\email{klee6@snapchat.com}
\authornotemark[2]
\affiliation{
  \institution{Snap Inc.}
  \country{United States of America}
}

\author{Chen Wang}
\email{chen.wang@snapchat.com}
\affiliation{
  \institution{Snap Inc.}
  \country{United States of America}
}

\renewcommand{\shortauthors}{Fu et al.}

\begin{abstract}
Collaborative filtering (CF) is one of the most successful and fundamental techniques in recommendation systems.
In recent years, Graph Neural Network (GNN)-based CF models, such as NGCF~\cite{wang2019neural}, LightGCN~\cite{he2020lightgcn} and GTN~\cite{fan2021graph} have achieved tremendous success
and significantly advanced the state-of-the-art.
While there is a rich literature of such works using advanced models for learning user and item representations separately, item recommendation is essentially a link prediction problem between users and items.
Furthermore, while there have been early works employing link prediction for collaborative filtering~\cite{chen2005link, chiluka2011link}, this trend has largely given way to works focused on aggregating information from user and item nodes, rather than modeling links directly.

In this paper, we propose a new linkage (connectivity) score for bipartite graphs, generalizing multiple standard link prediction methods.
We combine this new score with an iterative degree update process in the user-item interaction bipartite graph to exploit local graph structures without any node modeling. The result is a simple, non-deep learning model with only six learnable parameters.
Despite its simplicity, we demonstrate our approach significantly outperforms existing state-of-the-art GNN-based CF approaches on four widely used benchmarks.
In particular, on Amazon-Book, we demonstrate an over 60\% improvement for both Recall and NDCG.
We hope our work would invite the community to revisit the link prediction aspect of collaborative filtering, where significant performance gains could be achieved through aligning link prediction with item recommendations.
\end{abstract}



\maketitle

\section{Introduction}
In the Information Age we live in today, users are often confounded with the paradox of choice: how could one effectively find the best items to consume when there are just too many of them?
For decades, Recommendation Systems have been heavily researched to mitigate this issue, and often with great success in enhancing users' experiences~\cite{covington2016deep, ying2018graph, linden2003amazon}. 
Collaborative Filtering (CF) methods, which aim to utilize the explicit or implicit user-item interaction data within a service to find relevant items for users to consume,
are some of the most effective approaches widely adopted in the industry for  personalized recommendations~\cite{linden2003amazon, covington2016deep}.
Our work precisely focuses on this aspect of recommendation systems.

Within this domain, the widely accepted baseline has been to employ Matrix Factorization (MF)~\cite{koren2009matrix} to represent user and items in terms of latent factors, as achieved through either memory-based approaches~\cite{, hofmann2004latent, linden2003amazon} or model-based approaches~\cite{koren2009matrix, rendle2012bpr, rendle2021item}.
The core idea is to model users and items such that, ideally, similar users and items would have their representations located closely within an embedding space. The recommendation problem is then solved through matching users to items with the highest affinity, often determined through a dot product of their latent factors or with a neural network \cite{rendle2020neural}. Yet, these MF techniques have only made use of the user-item interaction data implicitly. When considering this data as a bipartite graph with users and items as the nodes, it is possible to explicitly incorporate such graph information for modeling: if a user has interacted with an item, then there exists a binary link between the two. In this regard, GNN-based CF models like NGCF \cite{wang2019neural}, LightGCN \cite{he2020lightgcn} and GTN \cite{fan2021graph} are the state-of-the-art models in exploiting such graph data for recommendations.

Item recommendation is essentially a link prediction problem on a user-item bipartite graph.
Despite the great success of employing GNN for collaborative filtering, a key limitation is that they still learn representations on nodes, and measure the affinity between two node representations to predict the presence of a link between the nodes, rather than modeling the link representation directly.
Before the deep learning era, Huang~et~al.~\cite{chen2005link} demonstrated that utilizing standard linkage scores (e.g., Common Neighbors, Preferential Attachment \cite{liben2007link}, or Katz Index~\cite{katz1953new}) outperforms vanilla user-based and item-based collaborative filtering on a book recommendation task.
Chiluka~et~al.~\cite{chiluka2011link} further showed that standard linkage scores are particularly more effective than CF on large-scale user-generated content, like YouTube and Flickr.
Recently, Zhang~et~al.~\cite{zhang2020revisiting} challenged the notion of using GNNs for link prediction, since its message-passing nature would lead to the same link representations for non-isomorphic links in the graph. To resolve this, they use a labeling trick to improve the link representation learning for link prediction, rather than just aggregating from two learned node representations. These works all point to the potential necessity to revisit link predictions for item recommendations.

In this paper, we propose a new linkage score for bipartite graphs, which generalizes several existing linkage scores like Common Neighbors~\cite{chen2005link}, Salton Cosine Similarity~\cite{salton1983introduction} and Leicht-Holme-Nerman Index~\cite{leicht2006vertex}. We then combine this linkage score with a lightweight iterative degree update step, which gives us a simple link prediction model with only six learnable parameters. Despite its simplicity, our proposed method significantly outperforms the state-of-the-art GNN-based CF models on multiple benchmark datasets. Notably, we achieve a larger than 60\% improvement on Amazon-Book compared to our nearest competitor. 
Apart from improved scores, our work is easy to implement and fast to train.

\subsection{Our Contributions}
The key contributions of our work could be summarized as below:
\begin{itemize}
    \item We demonstrate that our proposed method, a simple link prediction approach only exploiting the graph structure without any user/item modeling, can significantly outperform state-of-the-art GNN-based models with advanced user/item representation learning on several benchmark datasets.
    \item We propose \ourname{}, a new linkage score for link prediction on a bipartite graph, which generalizes multiple standard linkage scores in the literature. Our ablation study demonstrates this new learnable score function outperforms those standard linkage scores, which demonstrates the generality and superiority of our approach.
    \item We show that our method is robust to interaction noise commonly seen in real-world data, which makes it primed for use in an industrial setting.
\end{itemize}

\section{Background}
In this section, we describe and formulate the primary task we want to solve, and how related works have tackled the task in the past. We explain the issues surrounding existing approaches, which act as a bridge leading to our methodology in the next section to address these issues.

\subsection{Task Overview}
Our core task is recommending items through Collaborative Filtering (CF) with implicit feedback. Such feedback is implicit because it captures user behaviors (e.g. purchases and clicks), which implicitly indicates a user's interests, and is represented as the user-item interaction data. For every user, there exists at most one unique binary interaction with every item, and we ignore repeated interactions. Formally, let $\mathcal{U}$ and $\mathcal{I}$ be the sets of users and items respectively, and $\mathcal{O}$ be the observed interactions between some $u \in \mathcal{U}$ and $i \in \mathcal{I}$.  We define a binary interaction matrix $M \in \mathbb{B}^{\vert\mathcal{U}\vert \times \vert \mathcal{I}\vert}$ such that:

\begin{equation}
M_{jk} = 
    \begin{cases}
      1 & \text{if\ } (u_j, i_k) \in \mathcal{O} \\
      0 & \text{otherwise}
    \end{cases}
\end{equation}
Our goal is to exploit information from $M$ to learn a scoring function $\hy(u,i)$ that reflects the preference of a user $u \in \mathcal{U}$ for an item $i \in \mathcal{I}$. 
\begin{equation}
    \hy(u,i) = \hyu{i} ,\quad \hy : \mathcal{U} \times \mathcal{I} \rightarrow \mathbb{R}
\end{equation}
The scoring function $\hy(u,i)$ is then used to sort the list of unseen items $\{ i \in \mathcal{I} \mid (u, i) \not\in \mathcal{O}\}$ for a given user $u$. The ranking of the top $k$ items per user can be evaluated through various metrics. In this paper, we follow previous works and evaluate the ranking by averaging $\text{Recall}@k$ and $\text{NDCG}@k$ across all users.

\subsection{Bipartite Graphs}
\label{sec:bipartite}
While the interaction data is represented as a matrix above, we can also represent the interaction data as a bipartite graph
\begin{equation}
  G = (\mathcal{U}, \mathcal{I}, \mathcal{O}),
\end{equation}
in which $\mathcal{U}$ and $\mathcal{I}$ are two disjoint sets of nodes and $\mathcal{O}$ is the set of undirected and unweighted edges connecting nodes in $\mathcal{U}$ to nodes in $\mathcal{I}$. This formulation will be relevant in later sections as we discuss link prediction based CF models, so we briefly review some properties of bipartite graphs.

The adjacency matrix $A$ for a bipartite graph takes the form 
\begin{equation}
   A = \left[ \begin{array}{cc}  0 &  M \\ 
      M^{\mathrm T} &  0 \end{array} \right]
  \label{eq:matrix-structure}
\end{equation}

To count the number of paths between nodes in $u \in \mathcal{U}$ and $i \in \mathcal{I}$ we can raise $A$ to an odd power.
\begin{equation}
  A^{2k+1} = \left[ \begin{array}{cc} 
       0 & ( M M^{\mathrm T})^k M  \\
      (M^{\mathrm T} M)^k M^{\mathrm T} & 0 
    \end{array} \right] 
\end{equation}

From here we can see that the number of paths of length $2k+1$ between a node $u$ and $i$ is given by
\begin{equation}
[(M M^{\mathrm T})^k M]_{ui}
\label{eq:path-lengths}
\end{equation}

Let $\Gamma(u_{i})$ be the set of neighbors for a user $u_{i} \in \mathcal{U}$, then $\Gamma(u_{i}) \cap u_{j} = \emptyset$ for any $u_{j} \in \mathcal{U}$, since our graph is bipartite so that $\Gamma(u_{i}) \subseteq \mathcal{I}$. Likewise, $\Gamma(u) \cap \Gamma(i) = \emptyset$ for any $u \in \mathcal{U}$ and $i \in \mathcal{I}$. We can define $\hg{u} = \Gamma(\Gamma(u))$ i.e. the neighbors of $u$'s neighbors. Therefore $\hg{u} \subseteq \mathcal{U}$ and $|\hg{u} \cap \Gamma(i)| \geq 0$. The same terms and statements apply to items in $\mathcal{I}$.

\subsection{Linkage Scores}\label{sec:linkage_scores}
We review classic linkage scores, which are used to measure the likelihood that a link should be formed between two nodes in a graph. We present augmented versions of these classic scores, which can be applied to bipartite graphs.

Recall we are interested in predicting links between nodes $u \in \mathcal{U}$ and $i \in \mathcal{I}$. A common term in the standard versions of the following scores is $|\Gamma(u) \cap \Gamma(i)|$. Where we can view $|\Gamma(u) \cap \Gamma(i)|$ as measuring the number of common neighbors, or equivalently, the number of length two paths between the two nodes. However, as described in Section~\ref{sec:bipartite} the two sets of nodes, neighbors of $u$ and neighbors of $i$, will always form the empty set under intersection. Therefore, we adjust the term to be $|\Gamma(u) \cap \hg{i}|$ following~\cite{chen2005link}, which is the number of paths of length three between the two nodes. Let $P(u,i)$ be the set of (item, user) tuples that connect $u$ to $i$ in our graph i.e. $P(u,i)=\{(i_x,u_x) : (u, i_x) \in \mathcal{O} \wedge (u_x, i_x) \in \mathcal{O} \wedge (u_x, i) \in \mathcal{O}\}$

\textbf{Common Neighbors (CN):} The CN score is widely used for link prediction due to its simplicity and effectiveness~\cite{newman2001clustering}. The standard version of CN measures the number of nodes that two nodes have both interacted with. Or in other words the number of paths of length two between two nodes. Therefore, in the bipartite version of CN we count the number of paths of length three between  two nodes. Thus, the score is 
\begin{equation}
\label{eqn:CN}
\begin{split}
\text{CN}(u,i)  &= |\Gamma(u)\cap\hg{i}| \\
                &= \sum_{(i_x,u_x) \in P(u,i)} 1 \\
                &= |P(u, i)|
\end{split}
\end{equation}

\textbf{Salton Cosine Similarity (SC):} SC~\cite{salton1983introduction} measures the cosine similarity between two nodes $u$ and $i$:
\begin{equation}
\label{eqn:SC}
\begin{split}
\text{SC}(u,i)  &=\frac{|\Gamma(u)\cap\hg{i}|}{\sqrt{|\Gamma(u)| \cdot |\Gamma(i)|}} \\
                &= \sum_{(i_x,u_x) \in P(u,i)} \frac{1}{\sqrt{d_{u} \cdot d_{i}}} \\
\end{split}
\end{equation}

\textbf{Leicht-Holme-Nerman (LHN):} LHN is similar to SC without the square root in the denominator and thus will shrink the score of high degree nodes more quickly~\cite{leicht2006vertex}.
\begin{equation}
\label{eqn:LHN}
\begin{split}
\text{LHN}(u,i)  &=\frac{|\Gamma(u)\cap\hg{i}|}{|\Gamma(u)| \cdot |\Gamma(i)|} \\
                &= \sum_{(i_x,u_x) \in P(u,i)} \frac{1}{d_{u} \cdot d_{i}} \\
\end{split}
\end{equation}

\textbf{Parameter-Dependent (PD):} Zhu~et~al.~\cite{zhu2012uncovering} proposed PD which includes an adjustable parameter $\lambda$ to recover the previous three linkage scores. Specifically, with $\lambda=0$ we recover CN Equation~\ref{eqn:CN}, with $\lambda=0.5$ we recover SC Equation~\ref{eqn:SC}, and with $\lambda=1$ we recover LHN Equation~\ref{eqn:LHN}. This idea is very similar to our proposed linkage score in Section~\ref{sec:linkprop}.

\begin{equation}
\begin{split}
\text{PD}(u,i)  &=\frac{|\Gamma(u)\cap\hg{i}|}{(|\Gamma(u)| \cdot |\Gamma(i)|)^\lambda} \\
                &= \sum_{(i_x,u_x) \in P(u,i)} \frac{1}{(d_{u} \cdot d_{i})^\lambda} \\
\end{split}
\end{equation}

\subsection{Label and Link Propagation}
Label propagation \cite{huang2020combining} is an established approach for semi-supervised learning in graphs, particularly for node classification. At its core, it assumes the presence of \textit{homophily} in the graph, where similar nodes tend to be connected together. The main goal is then to predict which class unlabeled nodes belong to, by propagating information from labeled nodes.

\begin{figure}
  \centering
  \includegraphics[width=0.45\textwidth]{{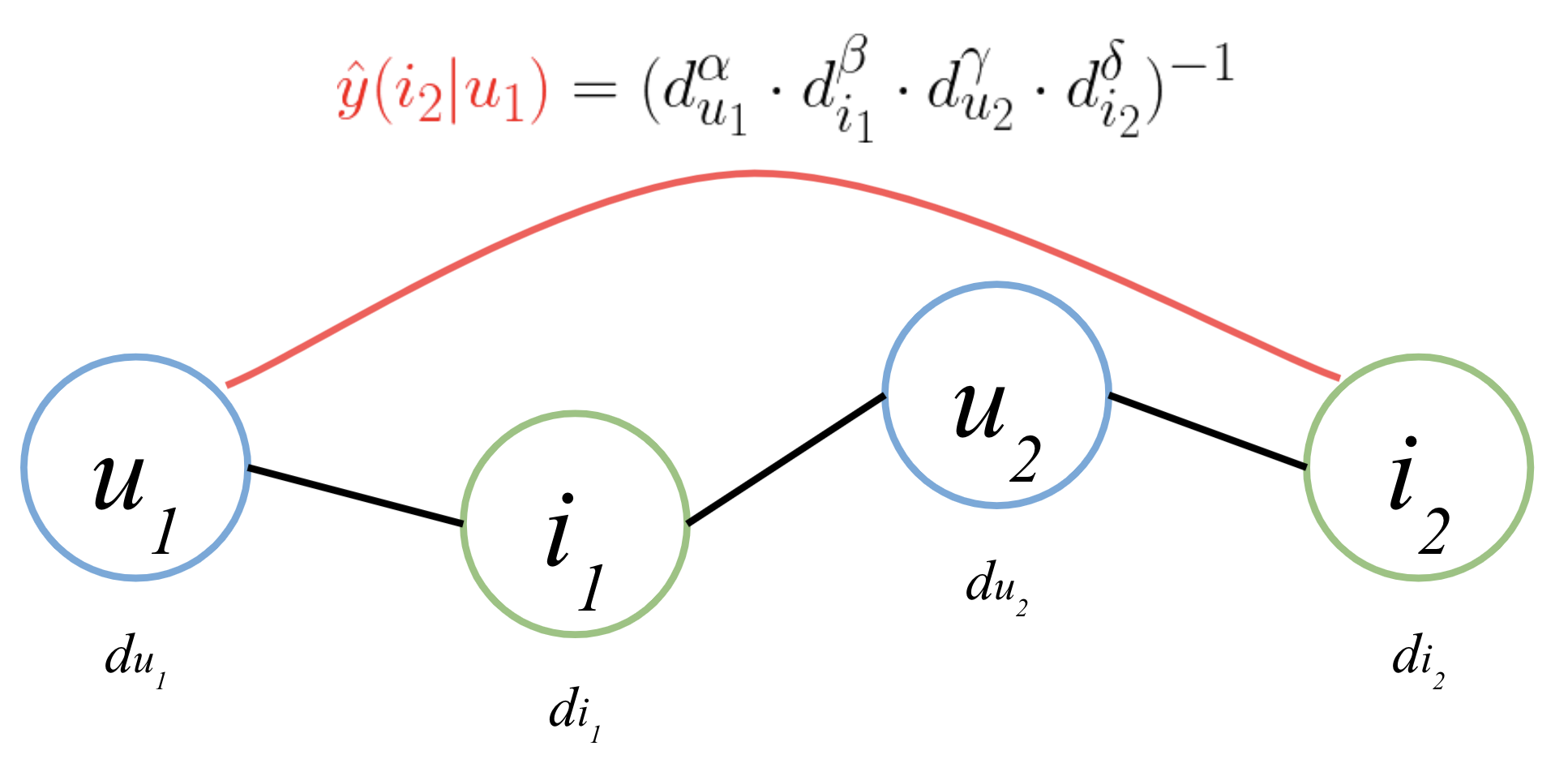}}
  \caption{ Illustration of \ourname{}.
  We define the linkage score $\hyuone{i_{2}}$ based on the observed $u_1 \leftrightarrow i_1 \leftrightarrow u_2 \leftrightarrow i_2$ path.
  }
  \label{fig:teaser}
\end{figure}

Our approach is closely related to label propagation in that we similarly propagate information from existing labeled data to unlabeled data, but with a core difference: we focus on links, and not nodes. 
There are no labels to assign to links, but rather the goal is to compute the strength (score) of a potential link.
Intuitively, this makes sense since the task of recommending items to users is inherently a link prediction one: the actual label assigned to users and items is secondary, and we are most concerned about whether there \textit{can} be a link or interaction between a user and item. Figure \ref{fig:teaser} illustrates an example of \ourname{}.
\section{Degree Aware Link Propagation}
In this section, we introduce our proposed link propagation methods \ourname{} and \ournamemulti{}. We first explain how \ourname{} exploits the user-item interaction graph to produce (soft) propagated links and how we compute the linkage scores for the predicted links. Next, we present \ournamemulti{}, which improves upon \ourname{} by performing multiple iterations. Finally, we explain our proposed training algorithm.

\subsection{Link Propagation}
\label{sec:linkprop}

Recall that our goal is to find highly possible connections between users and items in the interaction graph. The naive approach is to propagate links from existing, observed connections, such that a link is formed between a user $u$ and an item $i$ if the nodes are connected through an existing path in the interaction graph. For example, a link propagation could happen as follows for some $u_1, ..., u_k$ and $i_1, ..., i_k$: $u_1 \leftrightarrow i_1 \leftrightarrow u_2 \leftrightarrow i_2 \leftrightarrow ... \leftrightarrow u_k \leftrightarrow i_k$, where a link is denoted by $(.) \leftrightarrow (.)$. That is, a direct link is made directly between $u_1$ and $i_k$ from a path through their neighbors.

As mentioned in Section~\ref{sec:bipartite} any path connecting a user $u$ to an item $i$ must have length $2k+1$. In our method, we only consider the shortest valid path of length three for the following reasons. First, we note that LightGCN~\cite{he2020lightgcn} shows aggregating node embeddings from within three hops is effective and is the setting they use in their main experiments. Second, by using the same path length constraint as LightGCN~\cite{he2020lightgcn} we provide a more fair comparison to their method. Finally, if two nodes do not have any overlap in their local neighborhood even after a few hops, it is unlikely that they would share similar item preferences.

Naturally, it is overly simplistic to assume that every propagated link should have equal weight. Thus, our next step is to formulate a linkage score function $\hy$ to weigh a propagated link $u_{1} \leftrightarrow i_{2}$. We weigh the propagated link inverse proportionally to the degree of the nodes along the path connecting $u_{1}$ to $i_{2}$. Specifically, we score the link with the following equation:

\begin{equation}
\label{eqn:linkprop}
    \hyuone{i_{2}} = \sum_{(i_x,u_x) \in P(u_1,i_2)} \frac{1}{d_{u_1}^{\alpha} \cdot d_{i_x}^{\beta} \cdot d_{u_x}^{\gamma} \cdot d_{i_2}^{\delta}}
\end{equation}
where 
$P(u_1,i_2)=\{(i_x,u_x) : (u_1, i_x) \in \mathcal{O} \wedge (u_x, i_x) \in \mathcal{O} \wedge (u_x, i_2) \in \mathcal{O}\}$,
$d_{(\cdot)}$ is the node degree, and $\alpha, \beta, \gamma, \delta $ are learnable parameters. 

Intuitively, our scoring function attempts to weigh the likelihood of a link between any user-item based on the connectivity of the three-hops paths between them. Nodes on the path that are highly connected would have a high degree, which leads to a lower link weight, and vice versa. This is similar to the heuristic for TF-IDF \cite{rajaraman2011mining, jones1973index} where terms that appear in many documents tend to be less informative. Analogously, suppose for some $u_1 \leftrightarrow i_1 \leftrightarrow u_2 \leftrightarrow i_2$ interaction where some user $u_1$ and a low degree user $u_2$ enjoy the same obscure and low degree item $i_1$. Recommending user $u_1$ with some item $i_2$ that $u_2$ has also interacted with could prove to be informative due to its rarity, and a greater label weight is assigned to the $u_1 \leftrightarrow i_2$ link accordingly. 

From the proposed linkage score function in Equation~\ref{eqn:linkprop}, we can draw connections to several standard neighborhood-based link prediction score functions introduced in Sec.~\ref{sec:linkage_scores} by setting our parameters to specific values:
\begin{itemize}
    \item Common Neighbors (CN): $\alpha=\beta=\gamma=\delta = 0$
    \item Salton Cosine Similarity (SC): $\alpha = \delta = 0.5, \beta = \gamma = 0$
    \item Leicht-Holme-Nerman (LHN): $\alpha = \delta = 1, \beta = \gamma = 0$
    \item Parameter Dependent (PD): $\alpha = \delta = \lambda, \beta = \gamma = 0$
\end{itemize}
Thus, our model is able to learn the optimal values for $\alpha,\beta,\gamma,\delta$ which is more powerful and general than any one of these scoring functions. Our ablation study in Sec.~\ref{sec:ablation} shows making all of these parameters learnable is crucial to our strong performance.

We can further simplify our scoring function by noting that during evaluation, for each user, the score will be scaled by the same factor $d_{u_1}^{\alpha}$ i.e. 
\begin{equation}
    \hyuone{i_{2}} \propto \sum_{i_x,u_x \in P(u_1,i_2)} (d_{i_x}^{\beta} \cdot d_{u_x}^{\gamma} \cdot d_{i_2}^{\delta})^{-1}
\end{equation}
This is the final form of our scoring function for \ourname{}. We now have a continuous parameter space formed from $\beta$, $\gamma$ and $\delta$ alone, which are the three learnable parameters of \ourname{}.

Next, we derive the matrix version of \ourname{}. Let $d_{u_{i}}$ be the degree of user $u_{i}$ and $d_{\mathcal{U}} \in \mathbb{R}^{\vert\mathcal{U}\vert}$ be the vector of degrees for nodes in $\mathcal{U}$ where $[d_{\mathcal{U}}]_{i} = d_{u_{i}}$ and likewise for $d_{\mathcal{I}} \in \mathbb{R}^{\vert\mathcal{I}\vert}$. Let $d_{\mathcal{U}}^{\alpha}$ be the vector of degrees raised to $\alpha$ power, where $[d_{\mathcal{U}}^{\alpha}]_{i} = d_{u_{i}}^{\alpha}$

\begin{equation}
    D_{\alpha,\beta} =  d_{\mathcal{U}}^{-\alpha} \cdot (d_{\mathcal{I}}^{-\beta})^{\mathrm T} 
\end{equation}
\begin{equation}
    D_{\gamma,\delta} =  d_{\mathcal{U}}^{-\gamma} \cdot (d_{\mathcal{I}}^{-\delta})^{\mathrm T}
\end{equation}
Recall Equation~\ref{eq:path-lengths} for computing the number of odd path lengths in a bipartite graph, which is $M M^{\mathrm T} M$ for paths of length three. From here we can obtain our final matrix version of \ourname{}:

\begin{equation}
 \label{eqn:linkprop-matrix}
    L = (D_{\alpha,\beta} \odot M)M^{\mathrm T}(M \odot D_{\gamma,\delta})    
\end{equation}
where $\odot$ indicates the Hadamard product, and $L_{j,k} = \hat{y}(i_{k}|u_{j})$ as desired. In Algorithm~\ref{alg:code} we show the Numpy code for this method.

\subsection{Iterative Entity Degree Update}
Now we describe our \ournamemulti{} model, which leverages updated user/item degree values after an iteration of link propagation. Suppose we have performed an iteration of link propagation. Let $L^{(1)}$ be the link propagated interaction matrix from Equation~\ref{eqn:linkprop-matrix}, which contains both the observed and propagated links between users and items. We first mask out the observed links from $L^{(1)}$ and sort the propagated links based upon their score $\hy$. We then discard links that are not in the top $t$ proportion of links. For example, if we propagate 100 links and set $t=0.05$ then we will retain the five highest scoring links. We add the remaining links to the original interaction matrix and recompute the user/item degree values. We then use the updated user/item degree values and the \textit{original interaction matrix} as inputs to the next iteration of link propagation. Doing so, we can repeat the computation of $L^{(1)}$ for $r-1$ more iterations to obtain the final propagated matrix $L^{(r)}$.

However, this multiple iteration mechanism, $d_{u_1}$ now influences the propagated scores since we are sorting link scores across users, which means we need to re-introduce $d_{u_1}^{\alpha}$ in the computation of $L$. Thus, in total for this model variant, $t, r, \alpha$ are three new parameters we can learn, giving us six learnable parameters in total. For comparison, we refer to the model with $r=1$ as \ourname{}, and those with more than one propagation step as simply \ournamemulti{}.

\subsection{Model Training}
Given a continuous hypothesis set, one could use regular gradient descent based optimization techniques to find the optimal parameter combinations for our six learnable parameters. However, this typically requires a loss function (e.g. the BPR loss used in LightGCN), which acts as at most a \textit{proxy} for the final evaluation metric we care about. Furthermore, the optimal metric is dependent upon the overall system. For example, if the model is used during the retrieval stage rather than the ranking stage in a two-stage recommendation system \cite{covington2016deep}, then Recall could be a more appropriate metric compared to Normalized Discounted Cumulative Gain (NDCG)~\cite{liu2011learning}, as it does not consider the ordering of the retrieved items. In our work, we propose to directly optimize our model for NDCG without a proxy loss function. Since NDCG is non-differentiable, we quantize the parameter space and perform a coarse grid search for the optimal parameters. This is feasible because our model contains few parameters and we use a small search space for each parameter.

\subsection{Time Complexity Analysis}
\label{appendix:complexity}

In this section, we analyze and compare the time complexities of model training for both \ourname{} and LightGCN. Consider a dataset with $u$ users, $i$ items, and $l$ links, where $O(u) \approx O(i)$ and $O(l) \approx O(u^{1.5}) \approx O(i^{1.5})$.

During training of \ourname{}, we run $g$ inferences to complete a grid search for the best parameters. The dominating operation in terms of time complexity is calculating $MM^TM$, where matrix $M$ is a $u \times i$ matrix with $l$ non-zero values. With the sparse matrix $M$ stored in the Compressed Sparse Row (CSR) format, calculating $MM^TM$ takes $O(l \cdot min(u,i))$ time, where the selection between $u$ and $i$ is determined by calculating $MM^T$ or $M^TM$ first. Repeating this for $g$ times, the total time complexity of training \ourname{} is $O(gl \cdot min(u,i))$. Considering our assumptions for the relations between $O(u), O(i)$, and $O(l)$, we have $O(gl \cdot min(u,i)) \approx O(gu^{2.5})$.

As for training a LightGCN model, assume that it consists of $L$ Light Graph Convolutional layers and learns $d$ dimensional representations. Also, the model is trained for $e$ epochs with batch size $b$. Knowing that there are $l$ training samples with the BPR loss, there are $l/b$ batches in an epoch. Consequently, the total number of training steps is $e \cdot l/b$. For each training step, Light Graph Convolution with $O(ld)$ time complexity is conducted $L$ times, leading to a total complexity of $O(Lld)$. Combining all training steps, the overall time complexity of training a LightGCN model is $O(l^2Lde/b)$. Again, $O(l^2Lde/b) \approx O(u^3Lde/b)$.

Among all the factors, it is obvious that the total number of users, $u$, is significantly larger than any other variables in scale, and thus dominates the complexity. As a result, the training time complexity of \ourname{} is $O(gu^{2.5}) \approx O(u^{2.5})$, and that for LightGCN is $O(u^3Lde/b) \approx O(u^3)$.

\begin{algorithm}[t]
\caption{\ourname: Numpy Pseudocode}
\label{alg:code}
\algcomment{
\textbf{Notes}: \texttt{dot} is matrix multiplication. \texttt{M.T} is \texttt{M}'s transpose.
}
\definecolor{codeblue}{rgb}{0.25,0.5,0.5}
\definecolor{codekw}{rgb}{0.85, 0.18, 0.50}
\begin{lstlisting}[language=python]
# user_degrees: np array shape (U,) containing user degrees
# item_degrees: np array shape (I,) containing item degrees
# M: np array shape (U, I) containing interactions
# alpha, beta, gamma, delta: \ourname{} model parameters

# exponentiate degrees by model params
user_alpha = user_degrees**(-alpha)
item_beta = item_degrees**(-beta)
user_gamma = user_degrees**(-gamma)
item_delta = item_degrees**(-delta)

# outer products
alpha_beta = user_alpha.reshape((-1, 1)) * item_beta 
gamma_delta = user_gamma.reshape((-1, 1)) * item_delta

# hadamard products
M_alpha_beta = M * alpha_beta
M_gamma_delta = M * gamma_delta

L = M_alpha_beta.dot(M.T).dot(M_gamma_delta)

\end{lstlisting}
\end{algorithm}

\section{Discussion on LightGCN}
\label{analysis:lightgcn_connection}
Intuitively, LightGCN is trying to match the compatibility of some user and item via their learned node embeddings, followed by weighing this score proportionally to their (intermediate) node degrees. However, we argue that since item recommendations is intrinsically a \textit{link prediction task}, the biggest source of information could be obtained from the score weights for the potential link between two unconnected nodes, which we can further generalize to our proposed link propagation score $\hy$. If so, learning the exact node embeddings is secondary, and may be unnecessarily complicated as model training may not even converge perfectly. In fact, the general approach of learning node embeddings before computing the similarity between two embeddings could be seen as an approximation of a link between two nodes. Rather than taking this indirect approach, our method computes the link directly. In the next section, we demonstrate the superiority of our proposed link propagation method to prove this hypothesis.

\section{Experiments}

In this section, we first explain the settings we used for conducting fair and reproducible experiments. Next, we demonstrate the efficacy of our method against competitive state-of-the-art (SOTA) works. We also conduct ablation studies to understand our sources of improvements, and compare our proposed linkage score function to existing standard linkage scores. We further study how our approach is robust to varying levels of interaction noise that could be seen in real-world data. In total, we craft this section to answer the following Research Questions (RQ):
\begin{itemize}
    \item \textbf{RQ1:} How does our proposed method perform compared to competitive SOTA works?
    \item \textbf{RQ2:} How does our generalized linkage score function perform compared to standard linkage score function? What are the sources contributing most to our method's efficacy?
    \item \textbf{RQ3:} Is our method robust to interaction noise commonly seen in real-world data?
\end{itemize}

\subsection{Experimental Settings}

\begin{table}
\centering
\setlength{\tabcolsep}{2pt}
\caption{Basic statistics of benchmark datasets.}
\label{tab:dataset}
{
\begin{tabular}{|c||c|c|c|c|}
\hline
\multirow{2}{*}{\textbf{Datasets}} & \multicolumn{4}{c|}{\textbf{User-Item Interaction}}           \\ \cline{2-5} 
                                  & \textbf{\#Users} & \textbf{\#Items} & \textbf{\#Interactions} & \textbf{Sparsity\%} \\ \hline 
\textbf{Gowalla}                   & 29,858          & 40,981          & 1,027,370 & 99.92              \\ \hline
\textbf{Yelp2018}                  & 31,668          & 38,048          & 1,561,406 & 99.87              \\ \hline
\textbf{Amazon-Book}               & 52,643          & 91,599          & 2,984,108 & 99.94              \\ \hline
\textbf{LastFM}                    & 23,566           & 48,123           & 3,034,763 & 99.73               \\ \hline
\end{tabular}
}
\end{table}

\begin{table}
\centering
\setlength{\tabcolsep}{2pt}
\caption{Parameter values searched.}
\label{tab:parameter_search}
{
\begin{tabular}{|c||c|}
\hline
\textbf{Parameter} & \textbf{Values Searched} \\ \hline
$\alpha, \beta, \gamma, \delta$ & 0.0, 0.17, 0.34, 0.5, 0.67, 0.84, 1.0 \\ \hline
r & 1, 2, 3, 4 \\ \hline
t & 0.05, 0.1, 0.2, 0.3, 0.5, 1.0 \\ \hline
\end{tabular}
}
\end{table}

\begin{table*}
\centering
\caption{Learned Parameters.}
\label{tab:final_params}
{
\begin{tabular}{|c|c||c|c|c|c|c|c|}
\hline
\multirow{2}{*}{\textbf{Method}} & \multirow{2}{*}{\textbf{Dataset}} & \multicolumn{6}{c|}{\textbf{Parameters}} \\ \cline{3-8}
 & & $\alpha$ & $\beta$ & $\gamma$ & $\delta$ & $t$ & $r$ \\ \hline
 \multirow{4}{*}{\textbf{\ourname{}}} & \textbf{Gowalla}    & - & 0.5 & 0.67 & 0.34 & - & - \\ \cline{2-8}
 & \textbf{Yelp2018}                                        & - & 0.67 & 0.5 & 0.5 & - & - \\ \cline{2-8}
 & \textbf{Amazon-Book}                                     & - & 0.5 & 0.5 & 0.5 & - & - \\ \cline{2-8}
 & \textbf{LastFM}                                          & - & 0.67 & 0.67 & 0.34 & - & - \\ \hline
\multirow{4}{*}{\textbf{\ournamemulti{}}} & \textbf{Gowalla}    & 0.34 & 0.5 & 0.67 & 0.34 & 0.2 & 2 \\ \cline{2-8}
 & \textbf{Yelp2018}                                            & 0.34 & 0.67 & 0.5 & 0.5 & 0.05 & 3 \\ \cline{2-8}
 & \textbf{Amazon-Book}                                         & 0.34 & 0.5 & 0.5 & 0.5 & 0.1 & 3 \\ \cline{2-8}
 & \textbf{LastFM}                                              & 0.5 & 0.67 & 0.67 & 0.34 & 0.5 & 2 \\ \cline{2-8}
\hline
\end{tabular}
}
\end{table*}

\begin{table*}[htbp]
\centering
\setlength{\tabcolsep}{2pt}
\caption{The comparison of overall performance.}
\label{tab:comparsion_all}
\scalebox{1.0}{
\begin{tabular}{|c||c|c||c|c||c|c||c|c|}
\hline
\textbf{Datasets}              & \multicolumn{2}{c||}{\textbf{Gowalla}} & \multicolumn{2}{c||}{\textbf{Yelp2018}} & \multicolumn{2}{c||}{\textbf{Amazon-Book}} & \multicolumn{2}{c|}{\textbf{LastFM}}  \\ \hline
\textbf{Metrics}               & \textbf{Recall@20} & \textbf{NDCG@20} & \textbf{Recall@20}  & \textbf{NDCG@20} & \textbf{Recall@20}   & \textbf{NDCG@20}   & \textbf{Recall@20} & \textbf{NDCG@20} \\ \hline \hline
\textbf{MF~\cite{rendle2012bpr}}                         & 0.1299             & 0.111            & 0.0436              & 0.0353           & 0.0252               & 0.0198             & 0.0725             & 0.0614           \\ \hline 
\textbf{NeuCF~\cite{he2017neural}}                      & 0.1406             & 0.1211           & 0.045               & 0.0364           & 0.0259               & 0.0202             & 0.0723             & 0.0637           \\ \hline 
\textbf{GC-MC~\cite{berg2017graph}}                      & 0.1395             & 0.1204           & 0.0462              & 0.0379           & 0.0288               & 0.0224             & 0.0804             & 0.0736           \\ \hline 
\textbf{NGCF~\cite{wang2019neural}}                       & 0.156              & 0.1324           & 0.0581              & 0.0475           & 0.0338               & 0.0266             & 0.0774             & 0.0693           \\ \hline 
\textbf{Mult-VAE~\cite{liang2018variational}}                   & 0.1641             & 0.1335           & 0.0584              & 0.045            & 0.0407               & 0.0315             & 0.078              & 0.07             \\ \hline 
\textbf{DGCF~\cite{wang2020disentangled}}                       & 0.1794             & 0.1521           & 0.064               & 0.0522           & 0.0399               & 0.0308             & 0.0794             & 0.0748           \\ \hline 
\textbf{LightGCN~\cite{he2020lightgcn}}                   & 0.1823             & 0.1553           & 0.0649              & 0.0525           & 0.042                & 0.0327             & 0.085              & 0.076            \\ \hline 
\textbf{GTN~\cite{fan2021graph}}                        & 0.187              & \textbf{0.1588}           & 0.0679              & 0.0554           & 0.045                & 0.0346             & 0.0932             & 0.0857           \\ \hline \hline 
\textbf{Ours: \ourname{}}          & 0.1814             & 0.1477           & 0.0676              & 0.0559           & 0.0684               & 0.0559             & 0.1054             & 0.1025           \\ \hline 
\textbf{Ours: \ournamemulti{}}     & \textbf{0.1908}             & 0.1573           & \textbf{0.069}               & \textbf{0.0571}           & \textbf{0.0721}               & \textbf{0.0588}             & \textbf{0.1071}             & \textbf{0.1039}           \\ \hline
\textbf{Rel. Improvement (\%)}                & \textbf{2.03}               & -0.94             & \textbf{1.62}                & \textbf{3.07}             & \textbf{60.22}                 & \textbf{69.94}               & \textbf{14.91}               & \textbf{21.24}            \\ \hline
\end{tabular}
}
\end{table*}

We test our proposed models \ourname{} and \ournamemulti{} on four popular benchmark datasets: Gowalla~\cite{cho2011friendship}, Yelp-2018~\cite{yelp}, Amazon-book~\cite{ni2019justifying}, and LastFM~\cite{Bertin-Mahieux2011}. See Table~\ref{tab:dataset} for statistics summarizing these datasets.

For a fair comparison, we use the same preprocessed and split versions of these datasets as previous GNN-based methods~\cite{wang2019neural,he2020lightgcn,fan2021graph}, and we follow the same evaluation protocols and metrics. 
Specifically, we follow the setup in LightGCN where
only items the user has not previously interacted with are candidates for ranking, and evaluation is measured by computing the average $\text{Recall}@20$ and $\text{NDCG}@20$ across all users.

To prevent overfitting to the test dataset, we search for the optimal model parameters on a validation dataset, which we create by randomly sampling 10\% of a user's interacted items from the training data. Since the datasets were preprocessed to only include users with at least ten interacted items, we are guaranteed to have at least one item in the validation dataset for every user. 

Table~\ref{tab:parameter_search} shows the set of model parameters ($\alpha, \beta, \gamma, \delta, r, t$) that we search over when fitting our models. We can see that the total number of parameter and hyperparameter combinations searched is actually fairly small. For \ourname{} we search over $|\beta| * |\gamma| * |\delta| = 343$ combinations and for \ournamemulti{} we search over an additional $|\alpha| * |t| * |r| = 168$ values, which brings the total number of settings searched to $511$. Note in \ournamemulti{} we first find and fix the optimal $\beta, \gamma, \delta$ before searching over the additional values needed for \ournamemulti{}.

After we find the optimal model parameters on a validation dataset, we then perform inference using these settings from the observed links in the original training data. Our models directly output a relevance score for every unseen user-item pair. We use the predicted relevance scores to rank the unseen items for each user and compare our rankings to the test dataset.

In Table~\ref{tab:final_params} we show the values learned by \ourname{} and \ournamemulti{} on all four datasets.

\subsection{RQ1: Improving Item Recommendations}
\begin{table*}[htbp]
\centering
\caption{\textbf{Ablation on learnable parameters for \ourname.}}\label{tab:ablation}
\begin{tabular}{|ccc|c|c|c|c|c|}
\hline
\multicolumn{3}{|c|}{\textbf{Learnable}} & \textbf{Existing Standard} & \multicolumn{4}{|c|}{\textbf{Metrics}} \\ 
\multicolumn{3}{|c|}{\textbf{Parameter}} & \textbf{Linkage Score?} & \multicolumn{4}{|c|}{\textbf{Recall@20 / NDCG@20}} \\ \hline 
$\mathbf{\beta}$ & $\mathbf{\gamma}$ & $\mathbf{\delta}$ & & \textbf{Gowalla} & \textbf{Yelp2018} & \textbf{Amazon-Book} & \textbf{LastFM}  \\ \hline 
 0 & 0 & 1 & Leicht-Holme-Nerman (LHN)~\cite{leicht2006vertex} &                    0.0533 / 0.0360 & 0.0093 / 0.0075 & 0.0289 / 0.0219 & 0.0544 / 0.0432 \\ \hline 
 0 & 0 & 0.5 & Salton Cosine Similarity (SC)~\cite{salton1983introduction} &                    0.1252 / 0.0950 & 0.0553 / 0.0461 & 0.0506 / 0.0413 & 0.0936 / 0.0922 \\ \hline 
 0 & 0 & 0 & Common Neighbors (CN)~\cite{newman2001clustering} &                    0.1367 / 0.1142 & 0.0468 / 0.0385 & 0.0348 / 0.0278 & 0.0786 / 0.0752 \\ \hline 
\ding{52} & 0 & 0 & &           0.1597 / 0.1348 & 0.0513 / 0.0424 & 0.0403 / 0.0312 & 0.0904 / 0.0849 \\ \hline
 0 & \ding{52} & 0 & &           0.1548 / 0.1270 & 0.0496 / 0.0403 & 0.0440 / 0.0352 & 0.0845 / 0.0795 \\ \hline
 0 & 0 & \ding{52} & Parameter-Dependent (PD)~\cite{zhu2012uncovering} &           0.1397 / 0.1108 & 0.0554 / 0.0461 & 0.0506 / 0.0413 & 0.0937 / 0.0922 \\ \hline
\ding{52} & \ding{52} & 0 & &  \textbf{0.1849} / 0.1350 & 0.0568 / 0.0466 & 0.0532 / 0.0416 & 0.0986 / 0.0929 \\ \hline
 0 & \ding{52} & \ding{52} & &  0.1583 / 0.1251 & 0.0620 / 0.0514 & 0.0654 / 0.0540 & 0.0986 / 0.0954 \\ \hline
\ding{52} & 0 & \ding{52} & &  0.1615 / 0.1331 & 0.0584 / 0.0487 & 0.0527 / 0.0424 & 0.0999 / 0.0982 \\ \hline
\ding{52} & \ding{52} & \ding{52} & \ourname{} & 0.1814 / \textbf{0.1477} & \bf 0.0676 / 0.0559 & \bf 0.0684 / 0.0559 & \bf 0.1054 / 0.1025\\ 
\hline
\end{tabular}
\end{table*}

Our main results comparing to the state-of-the-art are shown in Table~\ref{tab:comparsion_all}. Despite the simplicity of our proposed models, \ournamemulti{} outperforms all other models on both metrics on all four datasets, except for $\text{NDCG}@20$ on Gowalla where our method performs slightly worse than GTN~\cite{fan2021graph}. Even \ourname{}, the simplified version with only one propagation iteration and no entity degree update, outperforms state-of-the-art models on three of the four datasets in terms of $\text{NDCG}@20$. This demonstrates the effectiveness of our proposed link propagation method over deep learning and classical MF methods.

It is noteworthy that our model outperforms other models on the Amazon-book dataset by an extremely large margin. This is because the number of users, items, and interactions in the Amazon-book dataset is much larger than the other three datasets. With more users, consequently more ranking lists, and more items needed to be satisfied by the model, the fixed-dimension latent space of users and items in node embedding based models may lack representational powers as the number of users and items scale. For example, the latent space dimension is fixed to 64 for all node embedding models. This may be enough for a model to learn informative embeddings for entities in the smaller Gowalla and Yelp2018 datasets, but is insufficient for a dataset as large as Amazon-book. This means that for such models, as the number of users, items, and interactions increase, they not only have to create a longer embedding lookup table for the nodes, but also requires a wider one with higher dimensions. 
It is then a challenge to apply these models in the real-world, where the scale of data is much larger than the datasets we used.
For clarity, we compare the performance of our model with a higher dimensional LightGCN in Appendix \ref{appendix:lightgcn_high_dim}, where we used the largest embedding dimensions without ``out of memory'' issue (16x larger), which increases LightGCN's performance by $16.4\%$ and $15.3\%$ on recall and NDCG. Although it's still far behind our method ($60.2\%$ and $69.9\%$ improvement).

In contrast, our model scales to growth in users/items much better. Without a need to explicitly learn a fixed-dimension embedding for each entity, it is not limited by the representation power of the latent space. Accordingly, we also removed the need for tuning the latent space dimension and dealing with the usual difficulties associated with training a gradient based model. This in turn vastly reduces the computational cost to train our model, which we provide more details in Section~\ref{appendix:complexity}.

\subsection{RQ2: Ablations}\label{sec:ablation}
In Table~\ref{tab:ablation} we show the performance of \ourname{} when excluding parameters from our linkage score defined in Equation~\ref{eqn:linkprop}. In order to exclude parameters, we fix the parameter value at zero. Note for all the different parameter combinations, we still search for the optimal parameter values using the train and validation datasets. 

From Table~\ref{tab:ablation} we can see that, as expected, excluding all parameters performs the worst except on Gowalla with respect to $\text{NDCG}@20$. Likewise, if we compare the single parameter models then $\delta$ for $i_2$ is the most crucial, except for on Gowalla. If we compare the performance of the models using only two parameters, we can see that each dataset varies on which combination provides the strongest result. Finally, we can see that using all three parameters provides the strongest results.

In addition, we also compare against existing standard linkage scores CN, SC, LHN and PD by fixing those parameters to specific values. Table~\ref{tab:ablation} also demonstrate that \ourname{} outperforms all of them by a large margin on all four datasets. This indicates the importance of making all of those parameters learnable.

Since we use the final metric to learn our parameters, we can easily change the metric to one required for our system. In Table~\ref{tab:recall} we show the additional gains in $\text{Recall}@20$ we are able to achieve by switching our metric from $\text{NDCG}@20$ to $\text{Recall}@20$.

\subsection{RQ3: Robustness to Noise}

\begin{figure*}[htbp]
 
\centering
{\subfigure[Gowalla - Recall@20]
{\includegraphics[width=0.245\linewidth]{{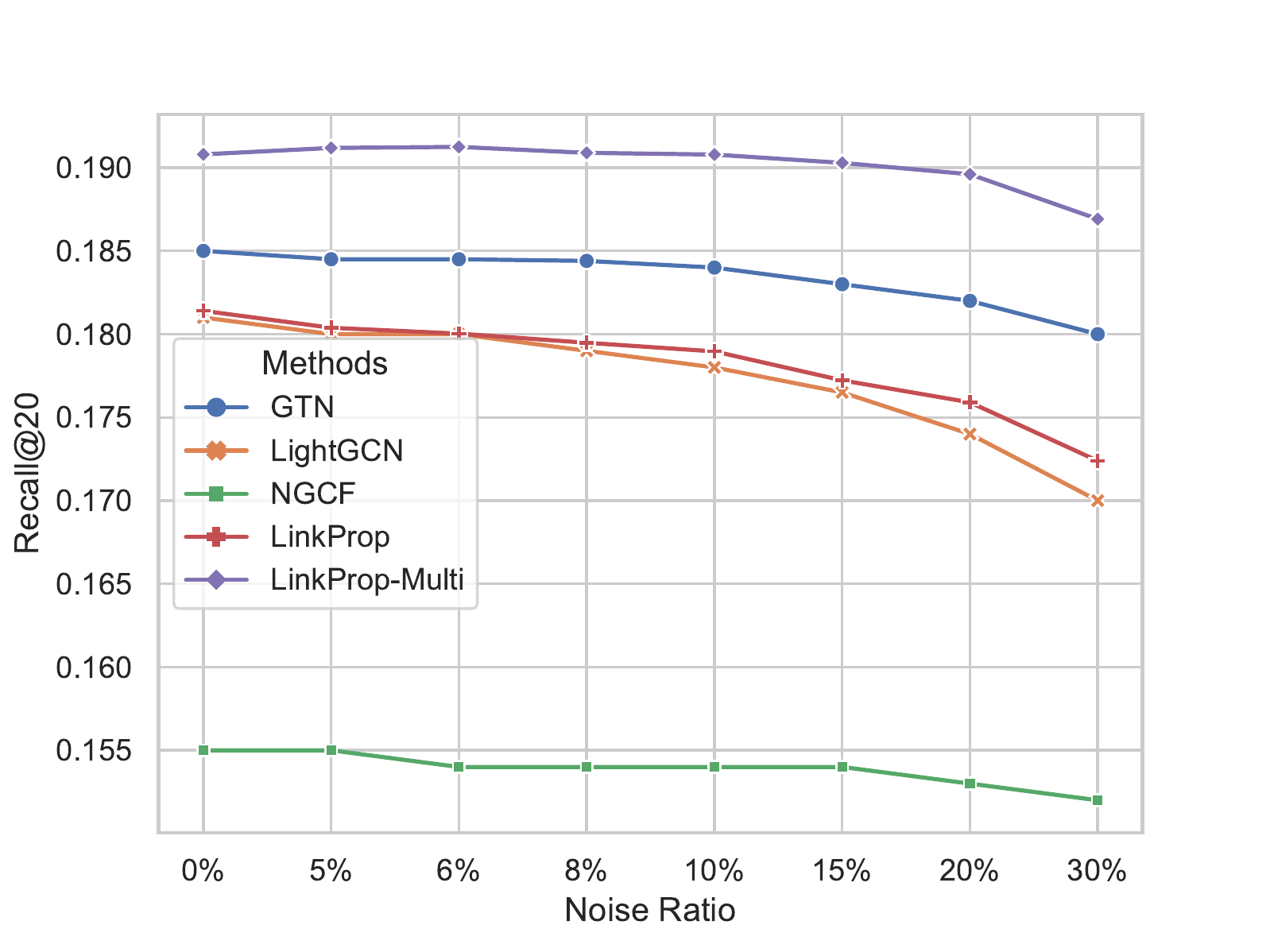}}}}
{\subfigure[Yelp2018 - Recall@20]
{\includegraphics[width=0.245\linewidth]{{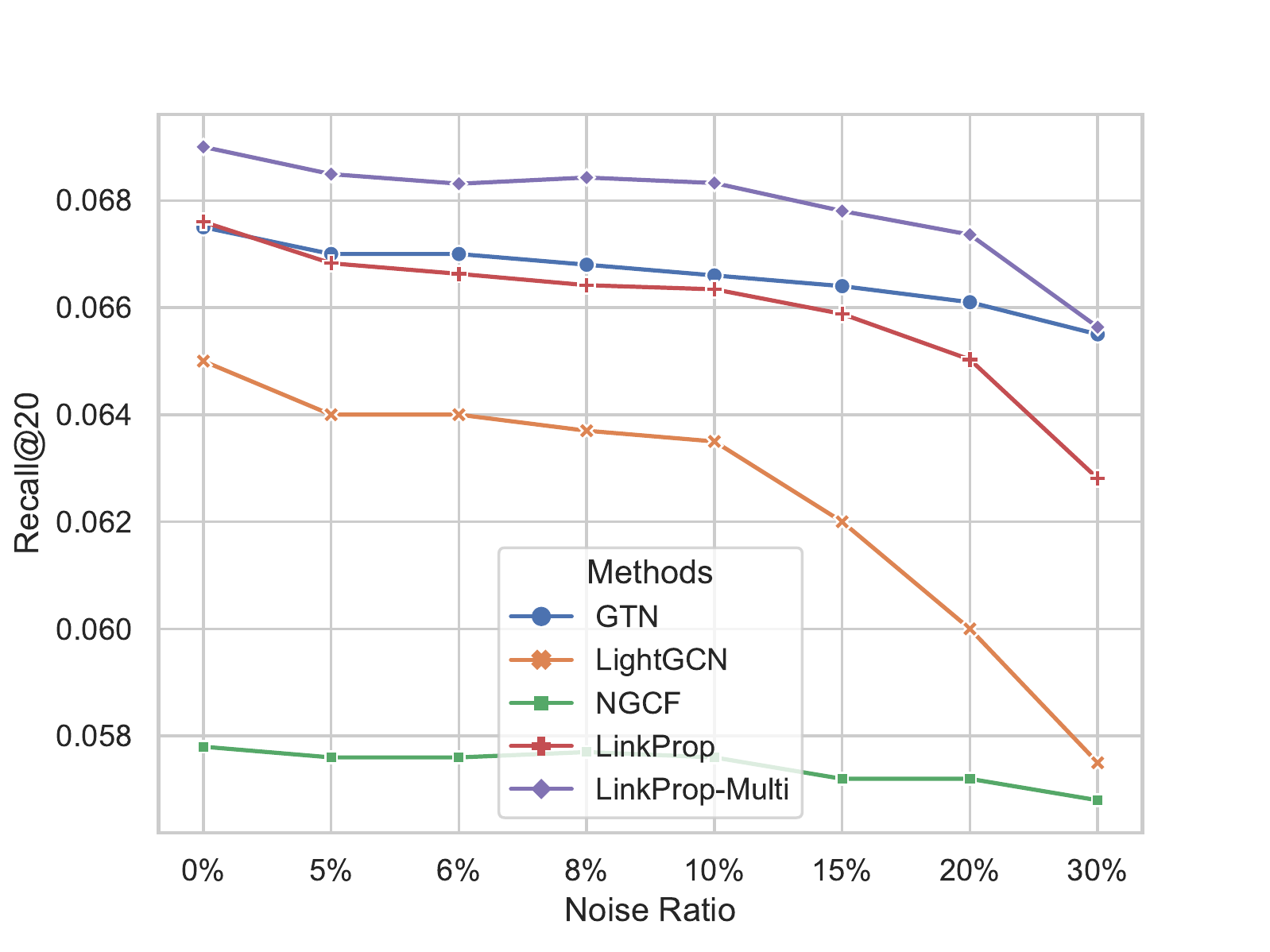}}}}
{\subfigure[Amazon-book - Recall@20]
{\includegraphics[width=0.245\linewidth]{{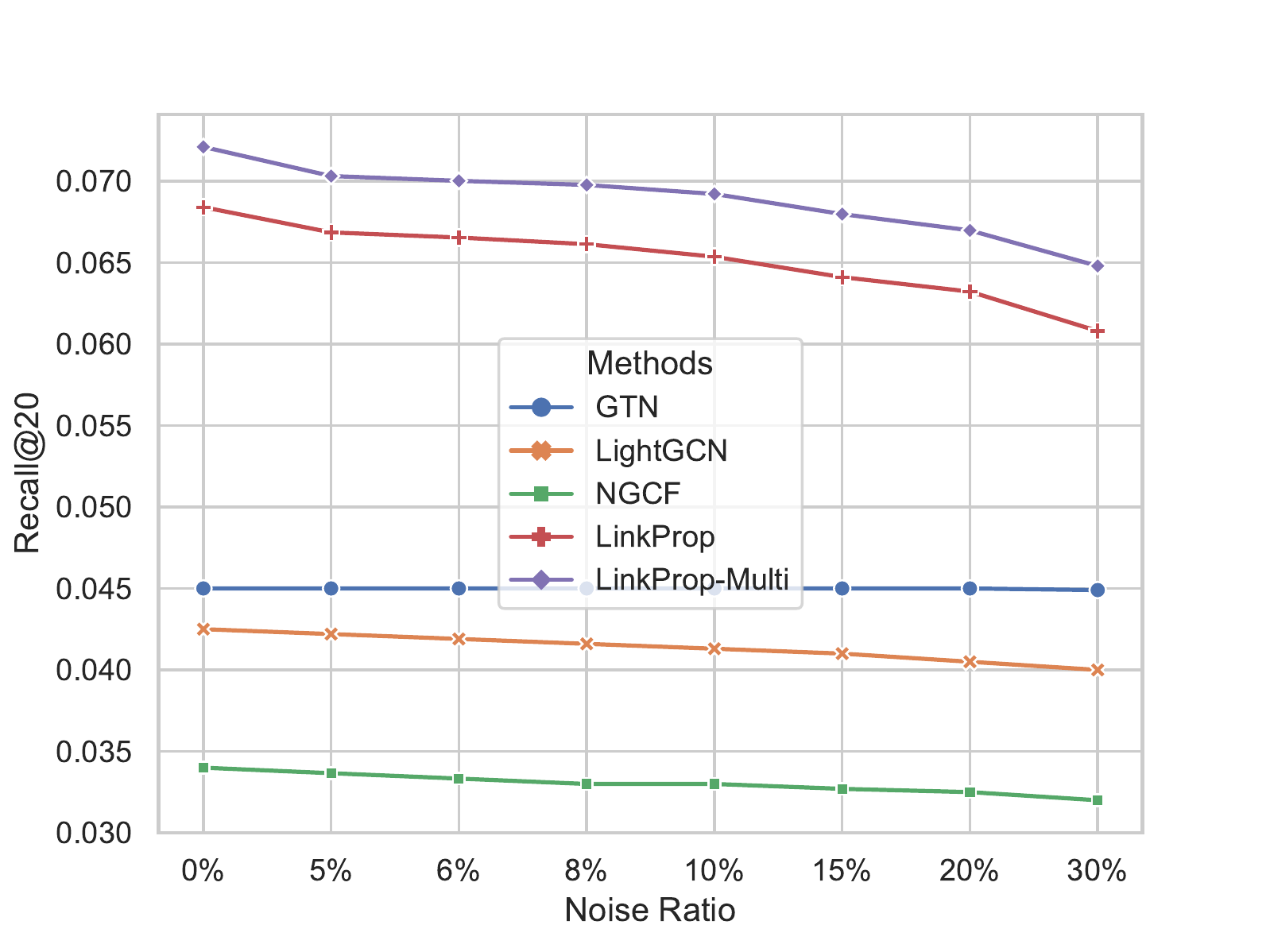}}}}
{\subfigure[LastFM - Recall@20]
{\includegraphics[width=0.245\linewidth]{{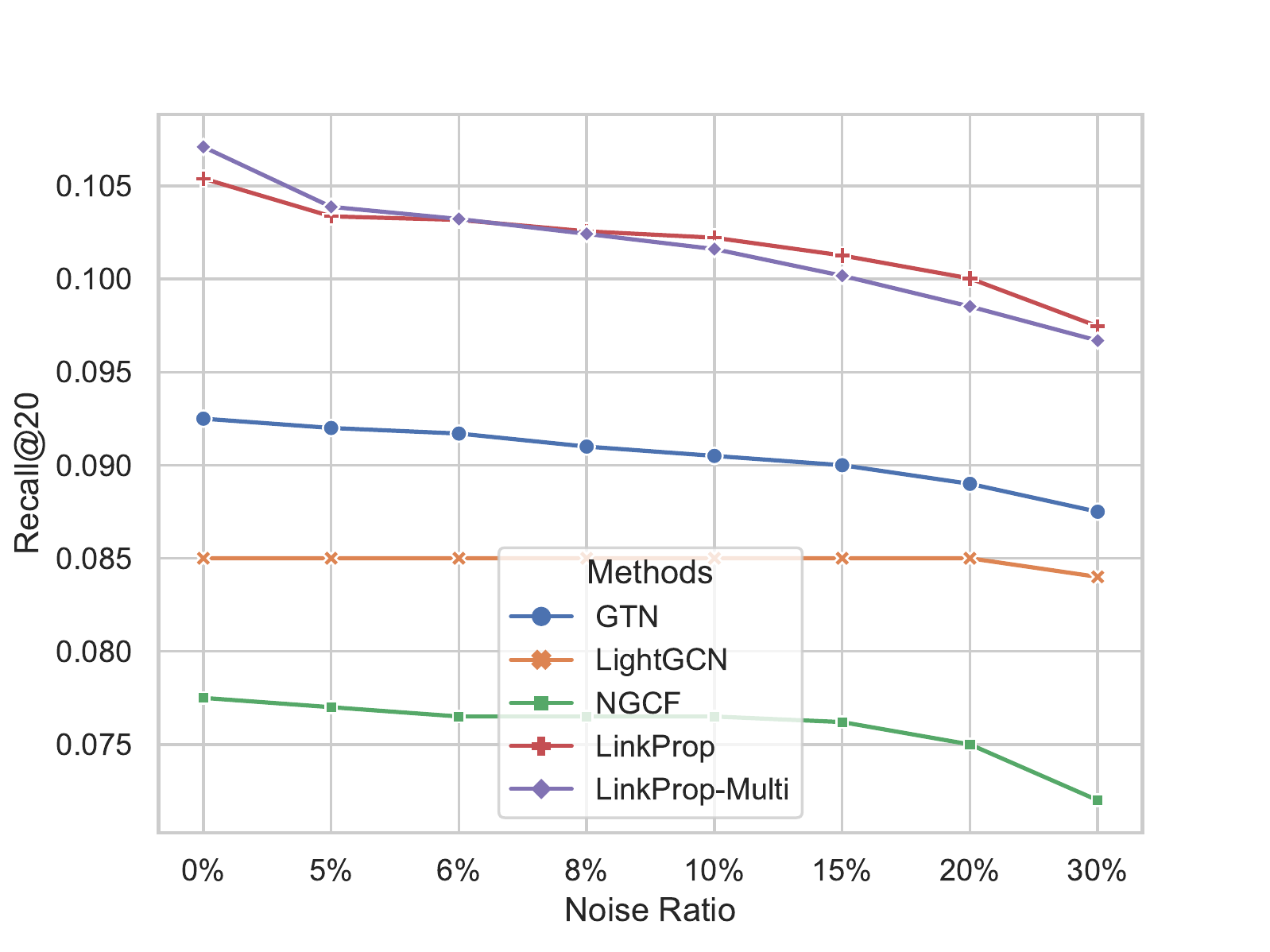}}}}

{\subfigure[Gowalla - NDCG@20]
{\includegraphics[width=0.245\linewidth]{{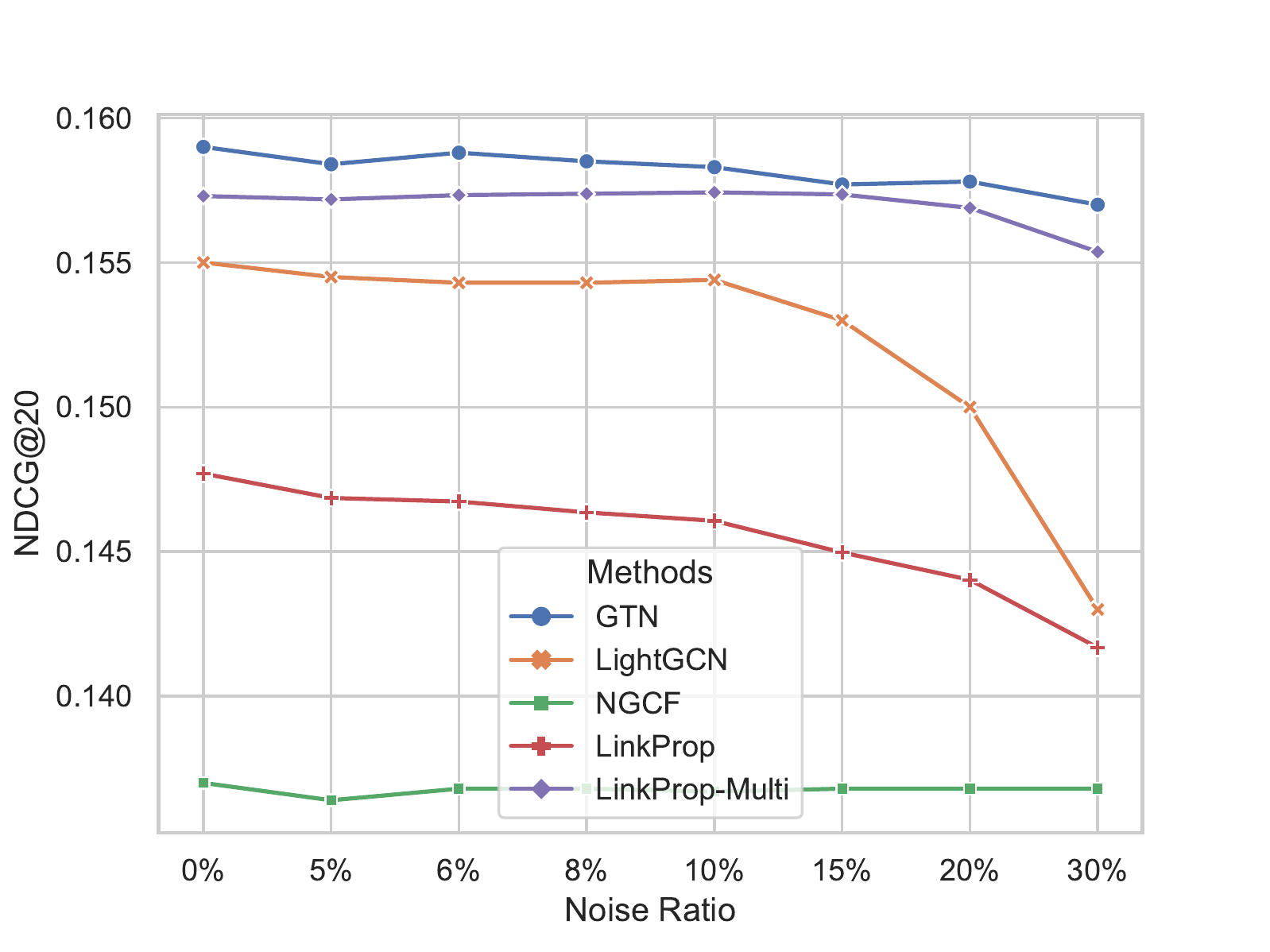}}}}
{\subfigure[Yelp2018 - NDCG@20]
{\includegraphics[width=0.245\linewidth]{{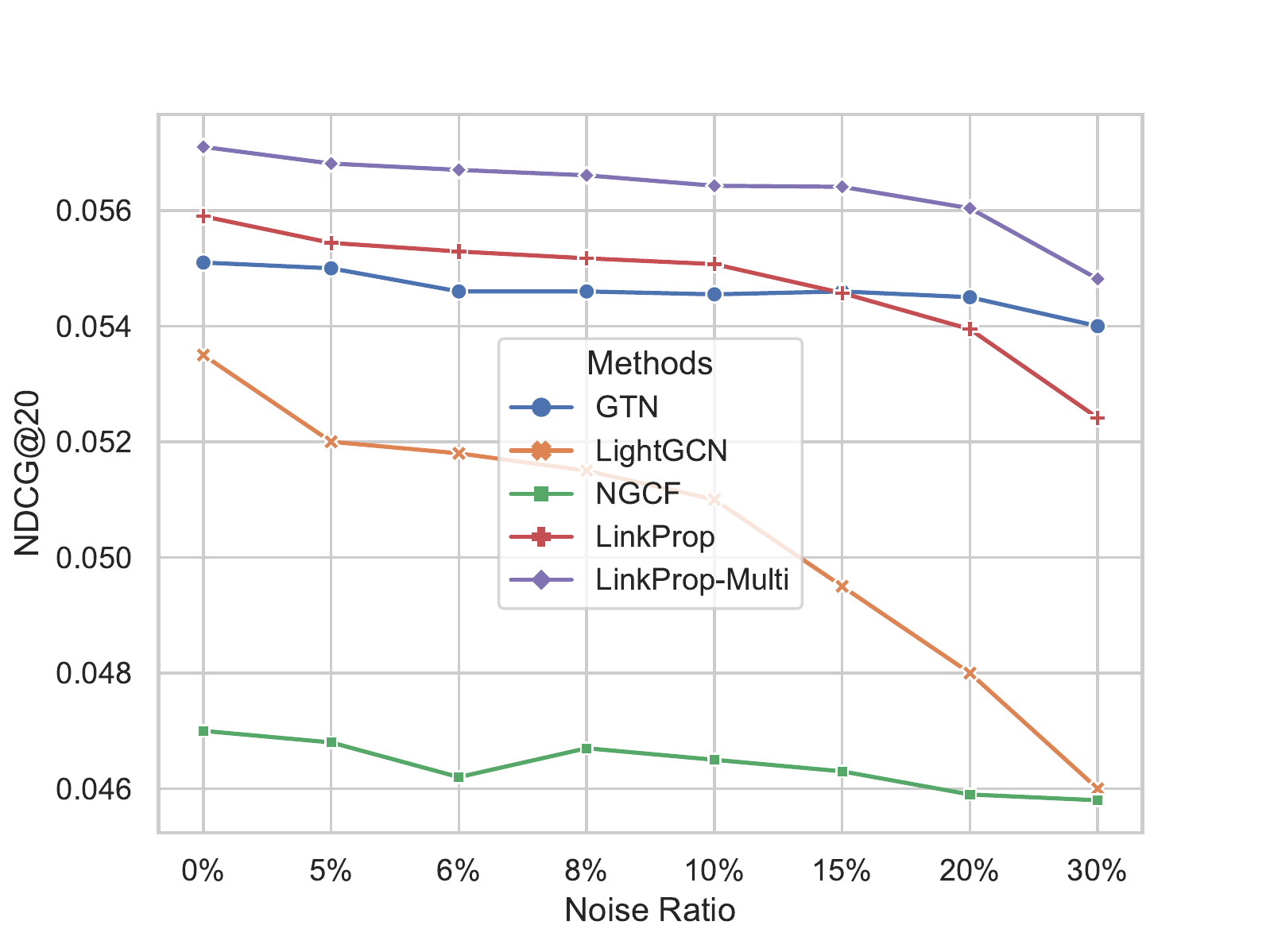}}}}
{\subfigure[Amazon-book - NDCG@20]
{\includegraphics[width=0.245\linewidth]{{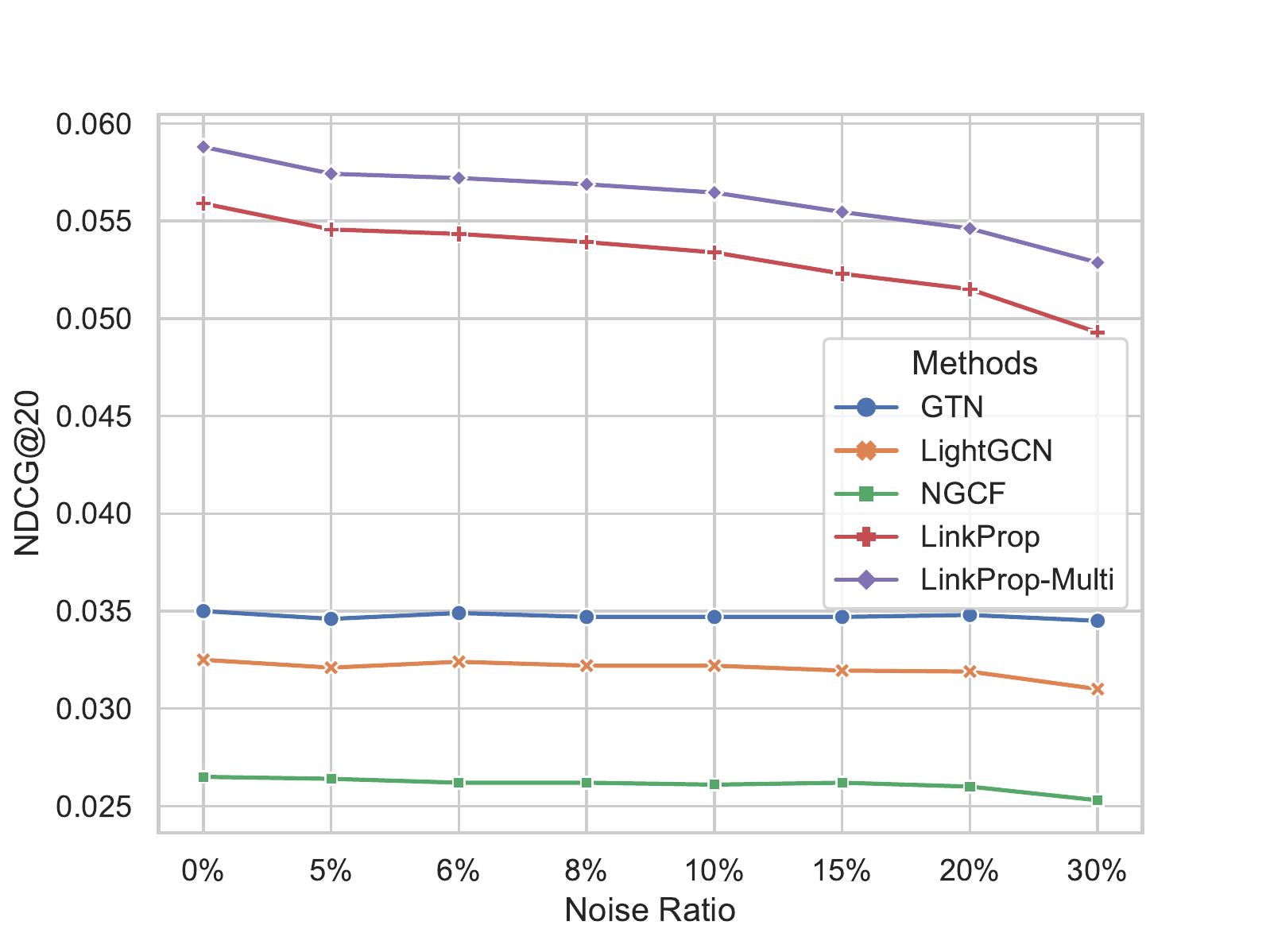}}}}
{\subfigure[LastFM - NDCG@20]
{\includegraphics[width=0.245\linewidth]{{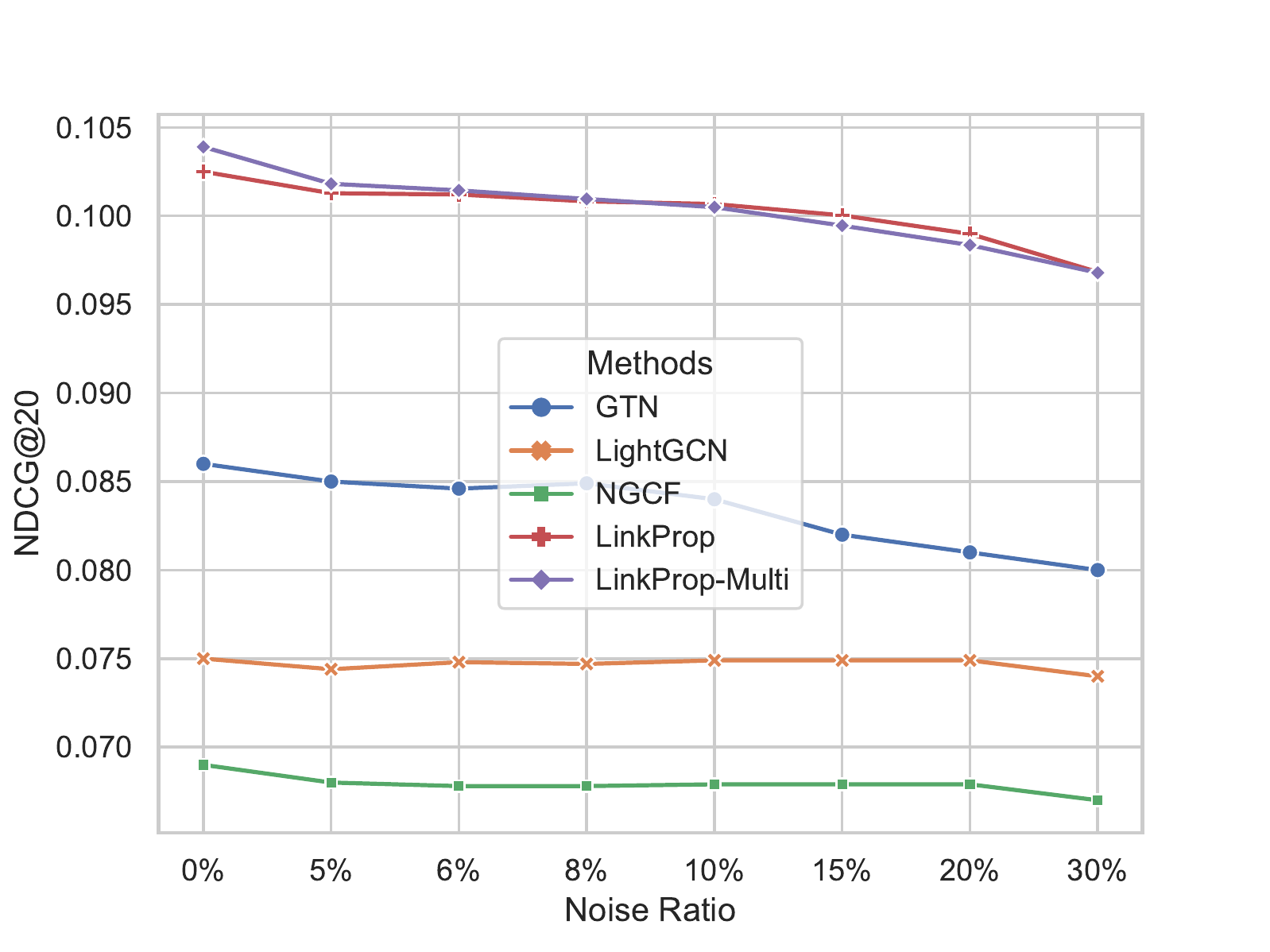}}}}
\caption{NDCG@20 and Recall@20 across different noise ratios perturbing the interaction graph.}
\label{fig:perturbation_rates}
\end{figure*}

We compare the robustness of our method against GNN-based CF models when noise is added to the interaction data. We follow the experimental settings proposed by GTN~\cite{fan2021graph} where we randomly insert fake interactions into the clean interaction graph such that the noise ratio $k$ as the percentage of the total interactions in the graph are fake.
As shown in Figure~\ref{fig:perturbation_rates} our methods maintain their effectiveness at a noise ratio of $30\%$ or less. Notably, compared to LightGCN, our method is much more robust across all noise ratios.
When comparing with GTN~\cite{fan2021graph}, it has the most robustness, although at a comparatively lower efficacy. However, this is expected since GTN is specifically designed to be robust to interaction noise.

\begin{table}
\centering
\setlength{\tabcolsep}{2pt}
\caption{Performance when optimizing for recall.}
\label{tab:recall}
\scalebox{1.0}{
\begin{tabular}{|c|c|c|}
\hline
\textbf{Datasets}              & \textbf{Gowalla} & \textbf{Yelp2018}  \\ \hline
\textbf{Metrics}               & \textbf{Recall@20} & \textbf{Recall@20}   \\ \hline \hline
\textbf{MF~\cite{rendle2012bpr}}                        & 0.1299             &  0.0436  \\ \hline 
\textbf{NeuCF~\cite{he2017neural}}                        & 0.1406  & 0.045   \\ \hline 
\textbf{GC-MC~\cite{berg2017graph}}                      & 0.1395 & 0.0462   \\ \hline 
\textbf{NGCF~\cite{wang2019neural}}                       & 0.156 & 0.0581    \\ \hline 
\textbf{Mult-VAE~\cite{liang2018variational}}                   & 0.1641 & 0.0584   \\ \hline 
\textbf{DGCF~\cite{wang2020disentangled}}                       & 0.1794 & 0.064   \\ \hline 
\textbf{LightGCN~\cite{he2020lightgcn}}                   & 0.1823 & 0.0649   \\ \hline 
\textbf{GTN~\cite{fan2021graph}}                        & 0.187 & 0.0679     \\ \hline 
\textbf{\ourname}                   & 0.1814 & 0.0679  \\ \hline 
\textbf{\ournamemulti}              & \textbf{0.1917} & \textbf{0.0700}  \\ \hline
\textbf{Rel. Improvement (\%)}                & \textbf{2.51} & \textbf{3.09}  \\ \hline
\end{tabular}
}
\end{table}

\section{Related Work}
Collaborative Filtering (CF) is a prevalent technique to build recommendation systems. It can be tackled with memory-based approaches~\cite{hofmann2004latent, linden2003amazon}, or more popularly, model-based approaches~\cite{koren2009matrix, rendle2012bpr, he2017neural, rendle2020neural, rendle2021item}. Recently, inspired by the success of Graph Convolutional Network (GCN)~\cite{kipf2016semi}, an explosion of GCN-based CF methods have been proposed: NGCF~\cite{wang2019neural}, GC-MC~\cite{berg2017graph}, PinSage~\cite{ying2018graph}, LightGCN~\cite{he2020lightgcn}, and GTN~\cite{fan2021graph}. However, a recent trend in GCN-based CF has been to build increasingly simple models that, surprisingly, perform better. LightGCN~\cite{he2020lightgcn} simplified the NGCF \cite{wang2019neural} approach by showing that the removal of nonlinearities and weight matrices led to improved results. On the path towards shallower models for recommendations, Rendle~et~al.~\cite{rendle2020neural} showed that the time-tested dot product method significantly outperforms a learned, through an MLP, embedding similarity function. The $EASE^R$ model~\cite{steck2019embarrassingly} avoided using any neural net modeling altogether, and simply used an item-to-item similarity matrix that discounts self-similarity to predict top items for each users. Similar to these works, we take a direction towards simplifying recommendation models, and end up with an approach that does not use any neural net modeling.

Our work is also closely related to the link prediction problem for graphs~\cite{liben2007link}. Specifically, we are solving the link prediction task on a bipartite graph. Huang~et~al.~\cite{chen2005link} reformulated standard linkage scores so that they can be applied to bipartite graphs. This reformulation consists of replacing terms that measure the number of common neighbors with a term for the number of three-hop paths. The adapted link measures were then applied to the recommendation task and brought gains over user and item based CF. Internal link prediction \cite{allali2011link} aims to predict potential connections between nodes in a bipartite graph by projecting the graph and adding links that don't alter the projected graph. Chiluka~et~al.~\cite{chiluka2011link} further showed that link prediction is particularly more effective than CF on large-scale user-generated content, like YouTube and Flickr. 
Our approach aims to use link prediction to address collaborative filtering.
As illustrated in Section~\ref{sec:linkprop}, through manipulating our parameters of $\alpha, \beta, \gamma, \delta$, we are able to recover many popular linkage scores such as the Salton Cosine Similarity~\cite{salton1983introduction}, Leicht-Holme-Nerman (LHN) score~\cite{leicht2006vertex}, and Parameter-Dependent (PD) score~\cite{zhu2012uncovering}. Amongst these scores, the PD-score is most similar to the \ourname{} score we proposed, but our score differs in that it further generalizes to include the neighbors between the nodes of the target link, and uses more than one parameter to control the individual degree terms. Furthermore, our eventual best performing \ournamemulti{} approach takes a further step in introducing a novel iterative entity degree update component.

\section{Conclusion}
In this paper, we introduced a simple and lightweight link propagation model for item recommendations, which significantly outperforms complex state-of-the-art models. Our method shifts away from the popular paradigm of creating complex (GNN-based) models which first learn user and item representations for finding matches, and instead opts to directly predict the existence of links. We argue this link prediction setup is the most natural one for item recommendations, which we support by showing even simple linkage score baselines beat SOTA GNN-based CF models. Towards this end, our method takes a step further to generalize many of these linkage score features into a new measure. Coupled with a novel iterative entity degree update component, our final \ournamemulti{}~method achieves the best performance across multiple benchmarks, including a significant margin of over 60\% improvement on Amazon-Book over our closest competitor. Furthermore, we show that such vast improvements incur only low computational complexity due to the simplicity of our model with only six learnable parameters. Finally, through showing how our work is closely connected to the GNN-based model, LightGCN, we offer a bridge to connect existing SOTA GNN-based models to the link prediction paradigm for item recommendations. We hope our work would inspire the community to revisit using link prediction for collaborative filtering and venture along this direction.

\clearpage
\bibliographystyle{ACM-Reference-Format}
\bibliography{main}

\clearpage
\appendix
\section{Appendix}

\subsection{Latent Dimension Tuning for LightGCN}
\label{appendix:lightgcn_high_dim}

As the embedding dimension of representation-based approaches are limited to 64 for the results in Table~\ref{tab:comparsion_all}, we suspect these models can perform better by increasing the embedding dimension. In order to compare \ournamemulti{} against the best performance possible for these models, we train larger LightGCN models by doubling their embedding dimension until we reach the memory limit of a single V100 GPU. The results for these larger models on Amazon-book are shown in Table~\ref{tab:higher_dim}.

Generally, LightGCN's performance improves as the embedding dimension increases. The largest dimension without a memory error is 1024, which achieves a 0.0489 $\text{Recall}@20$ and a 0.0377 $\text{NDCG}@20$, reflecting 16.4$\%$ and 15.3$\%$ improvements when compared to the original 64 dimension model. However, the performance is still far from that of \ournamemulti{}, which achieves a 0.0720 $\text{Recall}@20$ and a 0.0588 $\text{NDCG}@20$, a 47\% and 56\% relative improvement respectively.

From these results we can draw several conclusions. First, the optimal embedding dimension of representation-based approaches can be significantly higher than what is used in the literature, which could hurt their practicality for real world applications. Furthermore, for large datasets such as Amazon-book, the optimal embedding dimension is at least 1024, which is far higher than the original 64 dimension used in the original paper. Furthermore, in practice, the number of users and items can easily exceed those of Amazon-book, making scaling the model difficult. Second, even with a large dimension and improved performance, the approach is still significantly worse than our proposed \ournamemulti{}. This suggests that simply using larger models will not be enough for GNN-based CF methods to close the performance gap.


\begin{table}
\centering
\setlength{\tabcolsep}{2pt}
\caption{LightGCN performance with higher dimensionality. MemErr means a memory error is incurred on a V100 GPU.}
\label{tab:higher_dim}
{
\begin{tabular}{|c|c||c|c|}
\hline
\textbf{Model} & \textbf{dim} & \textbf{Recall@20} & \textbf{NDCG@20}\\ \hline
\multirow{6}{*}{\textbf{LightGCN}} & \textbf{64} & 0.0420 & 0.0327 \\ \cline{2-4}
 & \textbf{128} & 0.0460 & 0.0355 \\ \cline{2-4}
 & \textbf{256} & 0.0481 & 0.0371 \\ \cline{2-4}
 & \textbf{512} & 0.0487 & 0.0375 \\ \cline{2-4}
 & \textbf{1024} & 0.0489 & 0.0377 \\ \cline{2-4}
 & \textbf{2048} & MemErr & MemErr \\ \cline{2-4}
\hline
\textbf{\ournamemulti{}} & \textbf{-} & \textbf{0.0720} & \textbf{0.0588} \\ \hline
\end{tabular}
}
\end{table}

\end{document}